\documentclass[sigconf]{acmart}

\usepackage{booktabs}
\usepackage{cleveref}
\usepackage{dirtytalk}
\usepackage{natbib}
\usepackage{subfig}
\usepackage{enumitem}
\usepackage{url}
\usepackage{bbding} 
\usepackage{framed}

\setcopyright{rightsretained}

\copyrightyear{2017} 
\acmYear{2017} 
\setcopyright{acmcopyright}
\acmConference{EASE'17}{June 15-16, 2017}{Karlskrona, Sweden}\acmPrice{15.00}\acmDOI{http://dx.doi.org/10.1145/3084226.3084269}
\acmISBN{978-1-4503-4804-1/17/06}

\begin{document}

\title{The Links Between Agile Practices, Interpersonal Conflict, and Perceived Productivity}

\author{Lucas Gren}
\orcid{1234-5678-9012}
\affiliation{%
  \institution{Chalmers University of Technology and The University of Gothenburg\\}
  \streetaddress{The Department of Computer Science and Engineering}
  \city{Gothenburg} 
  \state{Sweden} 
  \postcode{412--92}
}
\email{lucas.gren@cse.gu.se}

\begin{abstract}
Agile processes explicitly focus more on team-work than more traditional management techniques when building software. With high velocity and responsiveness on team-level come the risk of interpersonal conflict in the agile organizations. Through a survey with 68 software developers from three large Swedish companies, I found that the presence of interpersonal conflict was negatively connected to the agile practices Iterative Development and Customer Access. The agile practices Iteration Planning and Iterative Development were positively linked to the measurement of the developers' perceived team productivity. However, Continuous Integration \& Testing was negatively connected to productivity. These results show which agile practices are directly linked to team productivity, but also, and more importantly, indicate which of the agile practices that might be more prone to not work as intended, when the team struggles with interpersonal conflict. Therefore, I argue that members of agile teams need training in conflict resolution techniques in order to lower the risk of interpersonal conflict negatively affecting team productivity.  
\end{abstract}


%
%

%
%



\maketitle

\section{Introduction}\label{sec:introduction}
The agile approach to software projects implies more focus on self-managing teams and group dynamics \citep{melnik}. With such focus, more psychological aspects like group norms and relationship conflicts, become increasingly more important to understand \citep{lenberg2015}. How group norms are set have been shown to increase performance in software engineering generally \citep{teh} as well as in agile software teams specifically \citep{stray2016exploring}. Group psychological aspects of teams have been shown to be key factors of successful agile teams \citep{grenjss2} and be utterly important to practitioners \citep{lenbergchase}. However, one key aspect of group dynamics, namely that of interpersonal conflict, has not been studied in the context of agile software development teams. 

In a study by \citet{liu2009negative} they also saw a negative effect of conflict on project success and these effects were not mediated by effective processes. However, their measurement of process included control over project costs, schedules, adherence to standards, etc., which implies a more plan-driven approach to projects. In a more recent and quite comprehensive study by \citet{nesterkin2016relationship}, they concluded that, in their partial mediation model, 60\% of the total effect of the relationship conflict and 80\% of the total effect of conflict management were mediated by team collaboration and goal-setting. Such results indicate that the team focus in agile software development is advantageous, however, we still know very little about how and what agile practices that are affected by interpersonal conflict. This study aims at filling parts of that gap and has therefore the following research questions:

\begin{itemize}
\item Which, if any, agile practices are positively or negatively associated with interpersonal conflict?
\item Which, if any, agile practices are positively or negatively associated with perceived productivity?
\end{itemize}

\section{Interpersonal Conflict and Software Engineering}
Traditionally in organizational psychology research, conflicts have been categorized into three main types; relation, process, and task. These categories simply refer to what the conflict is about, however, some scholars have suggested that the relationships between conflict types and performance are more complex \citep{behfar2008critical}. Relationship conflict have recently been shown to have indirect negative effects on both task-based and social aspects of team performance \citep{manata2016exploring}, which indicates that there are more complex relationships than a clear-cut separation between task-based and relational conflicts, as presented by for example \citet{trimmer2000information} in the software development domain and \citet{domino2003conflict} in the information systems domain. Within software engineering, an older study by \citet{gobeli1998managing} merely show that dysfunctional conflict management approaches have negative effects on results. 

In the broader research area of Information Systems Development an article from 2001 showed that, in the ISD context, the construct of interpersonal conflict (composed by disagreement, interference, and negative emotion) had less impact on project outcomes when good conflict management was in place \citep{barki2001interpersonal}, which was also shown by \citet{sawyer2001effects} in the same year. Within the requirements specification domain, interpersonal conflict was shown to be directly associated with requirements diversity, that, in turn, was negatively connected to project performance \citep{liu2011relationships}.  

\section{Interpersonal Conflict and Group Developmental Psychology}\label{inter}
\begin{figure}
\centerline{\includegraphics[width=90mm]{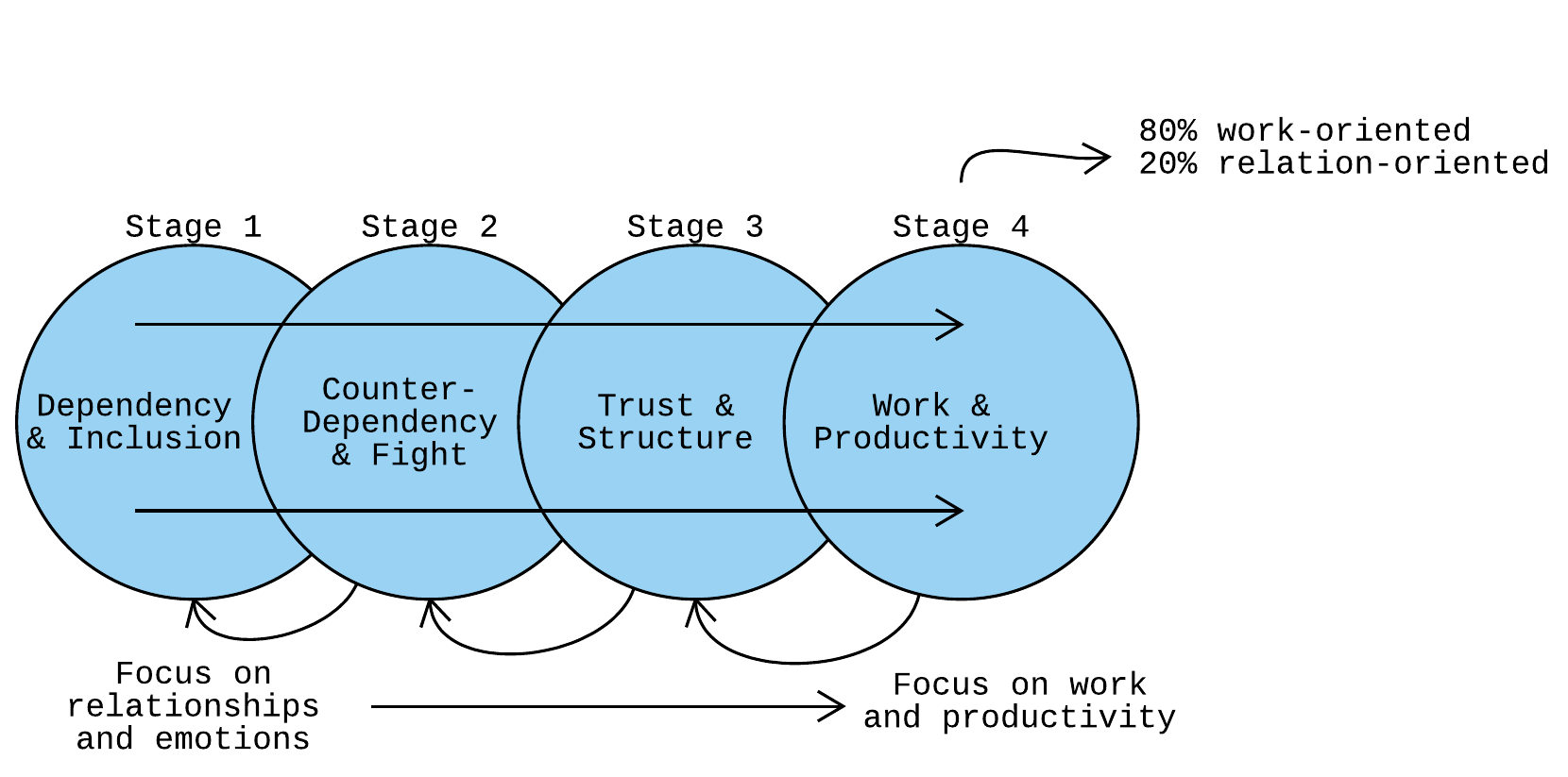}}
\caption{The Group Development Stages (adopted from \citet{wheelan2012})}
\label{fig:groupstages}
\end{figure}

The Integrated Model of Group Development (or IMGD) is a theory on group development that includes four different stages that all groups go through when moving towards becoming a high performing team \citep{wheelan}. These stages are illustrated in Figure~\ref{fig:groupstages} and describe overall patterns of which Scale 2 (Counter-Dependency and Fight) is the second stage of group development. No other group development measurement has been found that includes items regarding interpersonal conflict on its own scale (for a thorough review group processes research, see \citep{wheelandev}).

In the first stage (Dependency and Inclusion) the communication patterns are more polite since the group-members will be focused on safety and inclusion. The group members need to create a sense of belonging and also lay the foundation for how to interact within the group. Stage one (measured by Scale 1) is characterized by overly polite behavior, leader dependence, and very little conflict since the group-members first need to figure out how to relate to one another. The second stage is called Counter-Dependency and Fight, which means that the group is ready to start questioning goals, roles, and the structures of working together. During the second stage the group starts having conflict. These differences in opinion is a must in order to create clear roles based on competence and to make it possible to work together in a constructive way. The group members have to go through this more turbulent stage in order to build trust. Conflict is necessary in order to achieve shared perceptions of values, norms, and goals, which need to be set on group-level \citep{wheelan}. The important part is to turn the conflicts task- or process-related, and not get stuck in relationship conflicts \citep{jehn2001dynamic}. After this more turbulent questioning of how to work together, the group can focus more and more on finding roles, goals, and organize work in a more and more effective manner. A less mature division of work is measured in Stage 3 and what categorizes a high-performing and mature team is measured in Scale 4 \citep{wheelan}.

\citet{wheelan2012} was not the first researcher who categorized behavior into different stages of group development, but she contributed with a tool to measure these different stages with four scales by using a questionnaire. This tool has made it possible to measure and diagnose where a specific group is focusing its energy from a group developmental perspective (groups have been shown do more or less work in different stages over time \citep{wheelan2003}). The survey has a total of 60 items and provides a powerful tool for research on, and interventions in, teams. Scale 2 (GDQ2) is the ``Counter-Dependency and Fight'' and has been shown to correlate negatively with a set of effectiveness measures in different fields, for example,  groups that have high scores on GDQ2 finish projects slower~\citep{wheelan1998}, students perform worse on standardized test (SAT scores) if the faculty team scores high on GDQ2~\citep{wheelan1999}, and intensive care staff have higher death rates in surgery \citep{wheelan20032}. There are plenty group development models, but very few have been scientifically validated like the GDQ \citep{wheelan}. 

Furthermore, in a study by \citet{ocker2001relationship} in the Software Engineering domain, they showed that the level of group development was positively connected to the quality of the work product and the degree of satisfaction, which motivates using a group development measurement of conflict when studying software development teams.

\section{Method}\label{sec:method}
In this section I first present the participants, then the measured constructs, and finally how I conducted the data collection and analysis.

\subsection{Participants}
The data were collected from three Swedish large technology organizations and consisted of responses from 68 software developers. The first company was a multinational networking and telecommunications equipment and services company (with around 115,000 employees), the second company, was an aerospace and defense company (with around 14,000 employees), and the third company was an automotive parts manufacturing company (with around 160,000 employees). The teams consisted of 77 software developers in total, but 68 were present during the data collection (hence a response rate of 88\%). This high response rate was due to the fact that the surveys were filled out on paper and collected on site at a pre-scheduled time for each team. 

\subsection{Constructs}
Based on the research questions I needed to measure three different constructs in order to find answers. These are \emph{relationship conflict within team}, \emph{agile practices}, and \emph{perceived productivity}. The measurements for these three different constructs are described next.

\subsubsection{The Group Development Questionnaire (GDQ)}\label{sub:integratedgroup}
In order to measure relational interpersonal group conflict, I used a part of the Integrated Model of Group Development (or IMGD), namely Scale 2 that measures conflict related to Stage 2 of the group development model. All the items in the GDQ2 Scale can not be shared in this paper due to copyright reasons, however, I am allowed to include the three example items:

\begin{itemize}
\item People seem to have very different views about how things should be done in this group.
\item Members challenge the leader's ideas.
\item There is quite a bit of tension in the group at this time.
\end{itemize}

The question of not having formal leaders lead some participants to raise their hand and ask who the leader was in their agile team. Since all questionnaires were filled out on paper in the same room the researcher could provide the same clarification to all groups, namely to think of the leader as a person who takes initiative in the group, i.e., to see leadership as a function that can be shared in the team.

\subsubsection{The Perceptive Agile Measurement (PAM)}
The construct I used in order to measure agile practices and the behavior connected to these was the mature usage of eight agile practices as defined by \citet{so} and available in its entirety in their paper, however, the measured factors are:

\begin{itemize}
\item Iteration Planning
\item Iterative Development
\item Continuous Integration and Testing
\item Stand-Up Meetings
\item Customer Access
\item Customer Acceptance Tests
\item Retrospectives
\item Collocation
\end{itemize}

Due to all the different definitions and ambiguity of ``agility'' \citep{laanti2013definitions}, I chose this survey since it instead tries to capture the social-psychological behavior in connection to what the different practices try to achieve. It is also the only tool I have found that is validated through a factor analysis \citep{fabrigar} and a reliability analysis (using the Cronbach's $\alpha$ \citep{cronbach}) with a sample of $N=227$.

\subsubsection{Perceived Productivity}
In order to evaluate the effectiveness of the agile practices, I also asked the participants to rate their perceived productivity of their team. The participants were asked to rate their productivity using the single question ``In your opinion, how productive is this group?'' Measuring only developers self-assessed, and therefore only perceived, productivity is an open issue, however, \citet{graziotin2015feelings} argue that there is support from both psychology and software engineering studies to use perceived productivity as a proxy for objective productivity, since they are often tightly linked.

The group development measurement on Scale 2 was assessed on a 5-point Likert scale (1 = low agreement to the statement and 5 = high agreement). The agile items were assessed on a 7-point Likert scale (1 = never and 7 = always), with one exception being the Collocation items that were rated from 1 = the same room to 5 = different timezones. These scales were used for the simple reason that these measurements were developed and validated using these exact scales. The perceived productivity was rated from 1 (not productive at all) to 4 (very productive).

\subsection{Data collection and analysis}
The questionnaires were distributed in paper form and collected on site with all the teams present in the same room for the three companies separately, hence the high response rate (88\%). The researcher gave a short introduction to the research and stayed in the room to answer possible questions.

In order to investigate the connections between the three concepts I built two multiple linear regression models. In doing so, I wanted to see how much of the productivity and conflict measurements' variance I could predict by the agile practices' maturity. It is important to note the differences between predictive and causal models and in this study I only claim the former. 

\begin{figure}
\centerline{\includegraphics[scale=0.415]{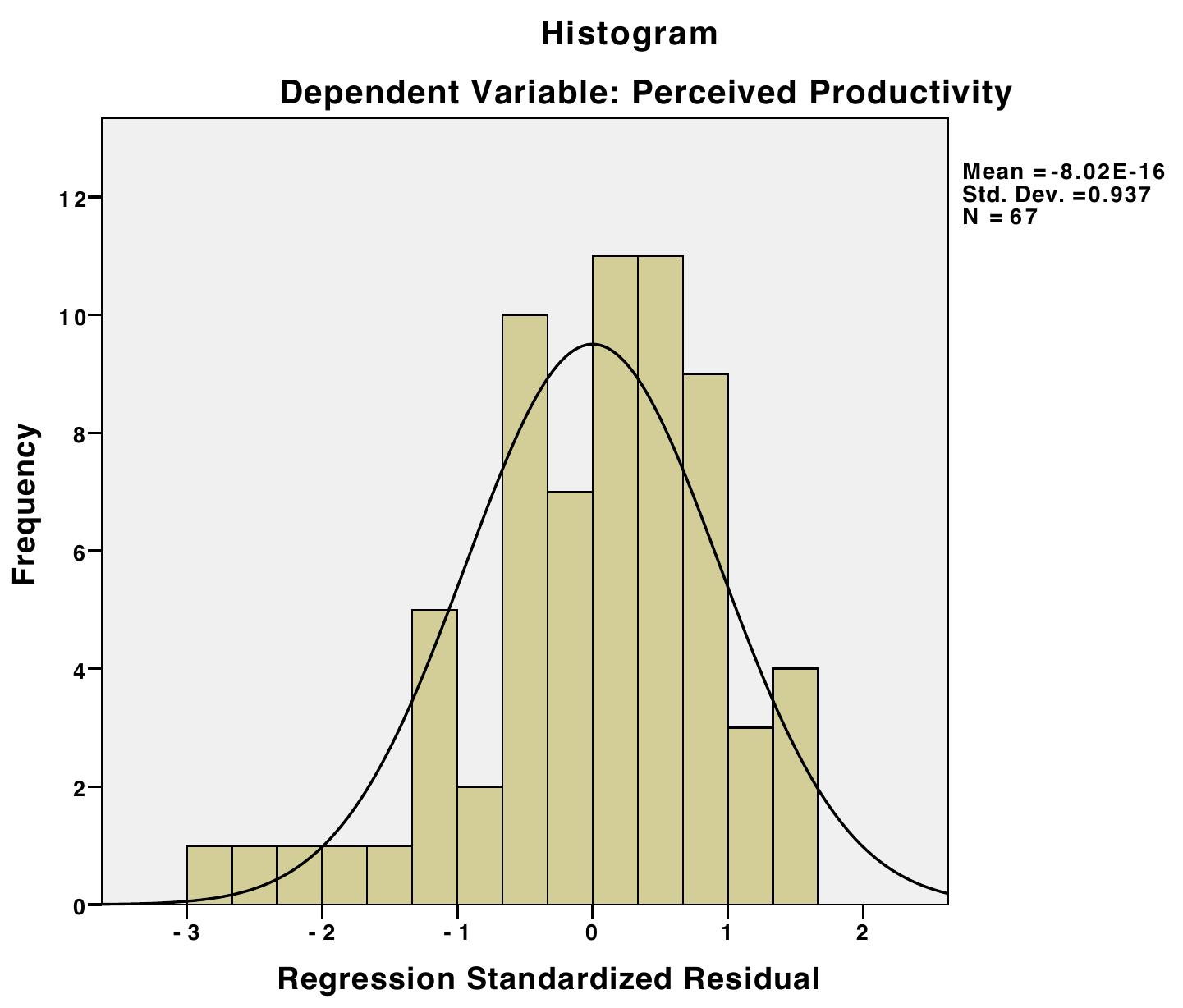}}
\caption{Frequency histogram with Productivity as Dependent Variable.}
\label{d}
\end{figure}

To evaluate if the data was normally distributed, I plotted frequency histograms for both multiple linear regression models. Figure~\ref{d} shows that the residuals are enough randomly scattered around the regression line but, in my second model, there might be an indication of a more complex relationship than linear between the ``Counter-Dependency and Fight (Scale 2)'' factor and the agile practices. Therefore, I proceeded and built a more complex model with the initially significant factors in order to obtain normally distributed residuals. I found a non-linear relationship between the agile practice Iteration Planning and Scale 2 and therefore suggest such a model in the results section. In order to assess the size of the effects in each analysis I calculated $\eta^2$ (often called $R^2$ in regression analysis) for each omnibus test (i.e.\ ANOVA) \citep{coheneffect}.

\section{Results}\label{sec:results}
The first ANOVA conducted with Scale 2 as dependent variable, gave us a significant omnibus test ($F=2.690, p=.014$), but the validation of the normality of the residuals showed clear deviations from normality.

\begin{figure}
\centerline{\includegraphics[scale=0.415]{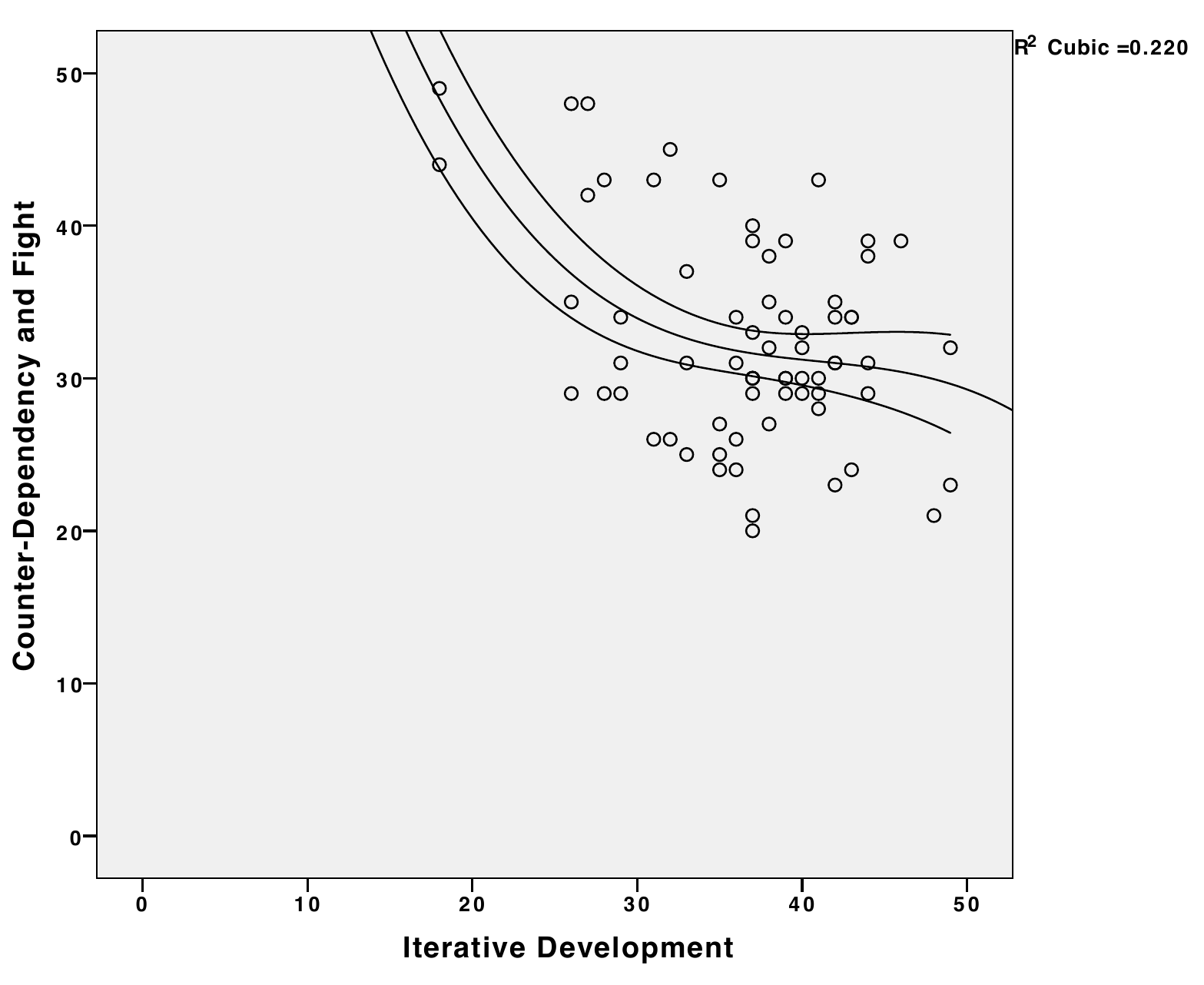}}
\caption{Scatter Dot Plot with fitted cubic regression model with a 95\% confidence interval.}
\label{scatter}
\end{figure}

In order to explore these relationships further, and without creating an overly complex model, I looked at scatter plots of all agile practices against the Scale 2 measurement. The significant factor Iterative Development from the first round showed clear non-linear relationship to Scale 2 as can be seen in Figure~\ref{scatter}.

\begin{figure}
\centerline{\includegraphics[scale=0.415]{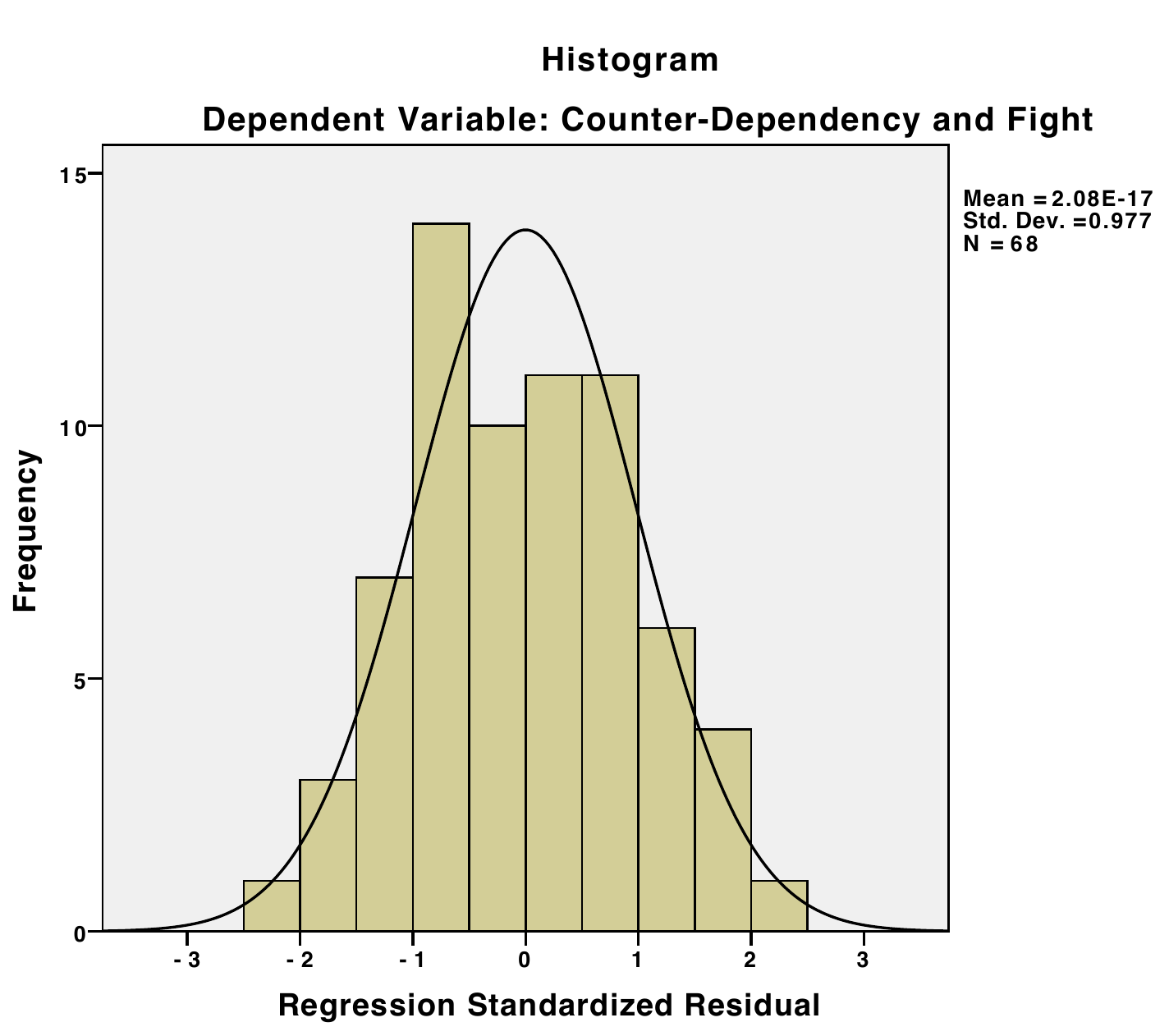}}
\caption{Frequency histogram for the second model with Scale 2 as Dependent Variable.}
\label{hist2}
\end{figure}

The second and more complex model gave the frequency histogram showed in Figure~\ref{hist2}. As can be seen, the standardized residuals are now nicely scattered around the zero, which gives us support for such a cubic relationship. These results revealed that the agile practice factors together could explain 26\% of the variance in the response variable GDQ2 ($F=7.623, p=.000$). Looking more closely at the significant factors we can see that the practices Iterative Development and Customer Access were the significant factors in the multiple linear regression, meaning that they were the ones significantly contributing to this explained variance (see Table~\ref{regscale2model2}). In an organizational research context such an effect is considered small, but still relevant \citep{coheneffect}, since explaining that much of the effect is difficult when researching complex systems that organizations also represent. 

\begin{table*}
\renewcommand{\arraystretch}{1.5}
\caption{Linear Regression Coefficients (Dependent Variable: GDQ2 with 68 valid cases).}
\label{regscale2model2}
\centering
\begin{tabular}{cccccccccccc}
\hline
\bfseries  Model & \bfseries Unstandardized B &\bfseries  Std. Error &\bfseries  Standardized B    &\bfseries  t  &\bfseries  $p$-value \\
\hline
(Constant) & 72.149 & 10.904  &  & 6.617 & .000* \\
\hline
Iterative Development & -1.393 &  .460 & -1.311 & -3.028 & .004*   \\
\hline
Customer Access & -.218 &  .096 & -.249 & -2.264 & .027*   \\
\hline
Iter.\ Dev.\ (Cubic) & .000 & .000 & .991 & 2.278 & .026*   \\
\hline
{*p$<$.05}  \\
\end{tabular}
\end{table*}

The ANOVA with Perceived Productivity as dependent variable revealed that the agile practices factors together could explain 44\% of the variance in the response variable Productivity ($F=5.713, p=.000$). Looking more closely at these significant factors we can see that the practices Iteration Planning and Iterative Development were the significant factors in the multiple linear regression, meaning that they were the ones significantly contributing to this explained variance (see Table~\ref{regprod}). However, the practice Continuous Integration and Testing was negatively associated to developers' perceived productivity, which means that low values on that measurement gave higher scores on the productivity measurement. All these three factors could together explain 44\% of the variance in the response and in an organizational research context such an effect size is considered medium \citep{coheneffect}. I will now discuss these results in more detail.

\begin{table*}
\renewcommand{\arraystretch}{1.5}
\caption{Linear Regression Coefficients (Dependent Variable: Productivity with 67 valid cases).}
\label{regprod}
\centering
\begin{tabular}{cccccccccccc}
\hline
\bfseries  Model & \bfseries Unstandardized B &\bfseries  Std. Error &\bfseries  Standardized B    &\bfseries  t  &\bfseries  $p$-value \\
\hline
(Constant) & 1.390 & .501 &  & 2.776 & .007* \\
\hline
Iteration Planning & .036 &  .011 & .452 & 3.408 & .001*   \\
\hline
Iterative Development & .048 &  .017 & .455 & 2.843 & .006*   \\
\hline
Cont. Int. \& Testing & -.025 &  .011 & -.346 & -2.169 & .034*   \\
\hline
Stand-Up Meetings & -.014 &  .020 & -.084 & -.700 &  .487  \\
\hline
Customer Access & .000 &  .011 & -.003 & -.026 & .980   \\
\hline
Cust. Accept. Tests & -.017 &  .010 & -.202 & -1.738 & .087   \\
\hline
Retrospectives & .018 &  .015 & .168 & 1.207 & .232   \\
\hline
Collocation & -.034 &  .018 & -.192 & -1.860 & .068   \\
\hline
{*p$<$.05}  \\
\end{tabular}
\end{table*}

\section{Discussion}\label{sec:discussion}
The results of this study indicate that, when the team struggles with interpersonal conflict, the agile practices Iterative Development and Customer Access are more prone to not work as intended. Also, the relationship between the agile practices and the conflict measurement were not linear, meaning, according to Figure~\ref{scatter}, that moderate conflict might have a larger effect as compared to little conflict, than moderate to extensive levels of conflict, i.e.\ the agile maturity of the practice Iterative Development decreases fast with quite little relational conflict introduced. The other measured practices showed no significant results in connection to the conflict measurement used. Looking more closely at the items included in the agile practices Iterative Development includes short iterations of code implementation, keeping deadlines, holding active discussions about prioritization with customers, delivering a potentially shippable product, meeting quality requirements of production code, and having working software as the primary measure of progress. It seems understandable that, when the team members are in a conflict stage, these activities become more difficult, which, in turn, will decrease productivity and effectiveness. 

Looking at the other significant factor, the practice Customer Access measured if the customer was reachable, if there were any bureaucratic hurdles in the communication, if the customer responded timely, and if the feedback from the customer was clear and clarified requirements or open issues to the developers. From a psychological perspective, people can easily notice if conflict is apparent when in contact with a team, which would naturally make us more careful in our communication. If group members have different views and have difficulty in agreeing on matters at hand, we would possibly get confused and not receive the inclusion and team spirit we would want as customers.

The results also show that the intended and mature use of the agile practices Iteration Planning and Iterative Development are connected to the developers' perceived team productivity. I have also showed that with higher scores on Continuous Integration and Testing came lower scores on this perceived productivity measurement. That means that the more continuous integration and testing the team conducts, the worse is the perceived team productivity. However, I do not have any external measurement of the productivity of the teams and can not draw conclusions on the actual productivity, and there are some empirical results indicating that more continuous integration implies higher productivity \citep{vasilescu2015quality}. To integrate continuously and to have rigorous testing might lower the perceived productivity, but, in fact, increase the external team productivity seen from an organizational perspective, i.e.\ writing more code feels more productive than working on re-factoring and testing code one has already written.

All in all, my results indicate the importance of having good tools to deal with conflict from a psychological perspective in order to achieve ``agility,'' i.e., the iterative development and customer relations needed to have that competitive advantage.

As a final remark, making employees aware of how conflicts work from a psychological and emotional perspective have already been shown effective, even in the ISD domain \citep{barki2001interpersonal}. My suggestion is that software engineering education should include negotiation and conflict resolution training, like the one presented by \citet{shell2001teaching}, especially in the agile context. I also believe having a formal structure for conflict resolution in agile software development organizations would increase productivity and job satisfaction. My study has provided empirical data on the importance of such approaches in order to leverage agile software development in the way it is intended. Two of the four statements in the agile manifesto \citep{fowler2001}, namely ``Individuals and interactions over processes and tools'' and ``Customer collaboration over contract negotiation'' are both connected to the results of this study. The agile manifesto is at the core of agile software development, and therefore, more research and guidelines of how to succeed with these in practice are much needed.

\section{Threats to Validity}\label{vts}
A limitation in this study is the operationalization of the two constructs used. The Perceptive Agile Measurement have been validated with 227 software engineers but the agility measurements have been shown difficult without taking context into account \citep{grenjss}. The Group Development Questionnaire have been thoroughly validated in its own field of organizational and social psychology, however, none of the validation studies were done in connection to software development. I also recognize the fact that Scale 2 of the GDQ might not cover all aspects of relational conflict, which means that this paper should only be seen as a first exploratory study of the connections between the two constructs. Further studies with higher resolution is therefore much needed in order to obtain knowledge of the more exact relationships between the two. 

I also acknowledge that using multiple linear regression analysis with a sample of 68 participants can be considered low with regards to how many variables I included in my questionnaire. However, conducting more advanced analyses, such as partial least squares path analysis, require larger sample size and were not used in this study due to the fact that I believe more qualitative data is needed first in order to know what associations to test (i.e., find more specific hypotheses). Since I lack knowledge of the internal and contextual relations between the agile practices, I did not want to run simple correlation analyses between all the categories, i.e., I wanted to see the predictive power of the agile practices in conjunction in relation to the interpersonal team conflict level.

As a side note, the more popular usage of more conservative nonparametric tests in software engineering research is often a good alternative in empirical research. However, when it comes to building regression models the assumption is that the residuals are normally distributed around the regression line. In my case, the first model I created broke this assumption, but the alternative is then to try to fit a more advance model to the data and then reevaluate the residual distribution. Since my second model showed normally distributed residuals around the regression line, even though curve-linear, the parametric assumption holds. The interpretability of such prediction models was considered very important in this initial study.

\section{Conclusions and future work}\label{sec:future}
This study set out to investigate which, if any, agile practices are negatively associated with interpersonal conflict, and which, if any, agile practices are, positively or negatively, associated with perceived productivity. Through conducting a survey and building two multiple linear regression models, I have found that the presence of interpersonal conflict was negatively connected to the agile practices Iterative Development and Customer Access. These findings are important contributions to the research and understanding of agile software development teams since it provides deeper understanding of the connections between intra-group conflict and the agile practices. While I have specifically focused on intra-group conflict (or interpersonal conflicts between group members), the connection between conflict and agile teams implies that my findings are likely to be of importance to both researcher and practitioners who try to understand and build agile teams. In terms of future research, I particularly suggest further replications that can offer higher resolutions of the connections between the constructs and more qualitative case studies explaining how teams can manage such conflict effectively.

\begin{acks}
I would like to thank Vard Antinyan for interesting discussions on this topic, possible explanations to the results, and feedback when writing this paper. 
\end{acks}

\bibliographystyle{ACM-Reference-Format}
\bibliography{references}  

\end{document}